\documentclass[twocolumn,aps,superscriptaddress,showpacs,nofootinbib,floatfix]{revtex4}
\usepackage{listings}
\usepackage[utf8]{inputenc}
\usepackage{hyperref}
\usepackage{amsmath}
\usepackage{amssymb}
\usepackage{amsfonts}
\usepackage{tabularx}
\usepackage{booktabs}
\usepackage{graphicx}
\usepackage{color}
\usepackage{multirow}
\usepackage[inline]{enumitem}
\usepackage[T2A,T1]{fontenc}
\graphicspath{{fig/}}
\definecolor{theblue}{RGB}{0,50,230}

\usepackage{hyperref}
\hypersetup{
  colorlinks=true,
  linkcolor=theblue,
  citecolor=theblue,
  urlcolor=theblue
}

\newcommand {\avg}[1]{\ensuremath{\langle\kern-1.0pt\langle#1\rangle\kern-1.0pt\rangle}}

\newlength\cmsFigWidth

\usepackage[normalem]{ulem}  

\renewcommand\sout{\bgroup \color{red} \ULdepth=-.5ex \ULset}

\begin{document}


\title{Pair production as a probe for the dynamics of nuclear fission and $\alpha$ decay}


\author{T. Settlemyre}
\affiliation{Cyclotron Institute, Texas A\&M University, College Station, Texas 77843, USA}
\author{H. Zheng}
\affiliation{School of Physics and Information Technology, Shaanxi Normal University, Xi'an 710119, China}
\author{A. Bonasera}
\affiliation{Cyclotron Institute, Texas A\&M University, College Station, Texas 77843, USA}
\affiliation{Laboratori Nazionali del Sud, INFN, via Santa Sofia, 62, 95123 Catania, Italy}


\begin{abstract}
Electron-positron pairs can be produced via the Schwinger mechanism in the presence of strong electric fields. In particular, the fields involved in $\alpha$ decay and nuclear fission are strong enough to produce them. The energy of the $e^+e^-$ pair is related to the relative distance and velocity of the daughter nuclei. Thus, the energy distribution of the produced pairs can give information about the dynamics of the fission and $\alpha$ decay processes. A neck model of nuclear fission is used to illustrate how the pairs can be used as a probe of the dynamics.
\end{abstract}

\maketitle

The Coulomb force is mediated by the exchange of a virtual photon, which may result in the creation and annihilation of a virtual $e^+e^-$ pair. This higher order effect causes vacuum polarization. The resulting correction to the Coulomb potential was found by Uehling \cite{Uehling:1935uj}. This correction is important in $p$-$p$ scattering \cite{prestonbook} and in nuclear scattering systems more generally \cite{Settlemyre:2021tbe}, where it is on the order of one percent of the Coulomb energy. Near the Coulomb barrier, the nuclear and Coulomb energies roughly cancel, so a small change in the Coulomb energy could significantly affect the height of the Coulomb barrier. A change in the height of the Coulomb barrier will affect the cross section of sub-barrier fusion \cite{assenbaum}, e.g. Carbon-Carbon fusion in the cores of stars.   

When the available energy exceeds twice electron mass, real $e^+e^-$ pairs can be created during the dynamics in the presence of strong fields  \cite{Schwinger:1951nm, wongbook, Blaschke:2019pnj, carrollbook, Voskresensky:2021okp, popov}. 
This effect may increase the fusion cross section above the adiabatic limit \cite{pairproduction, Kimura:2004dw, Krauss:1987onj, Engstler:1988tfw, Shoppa:1993zz, Musumarra:2001xd, Kimura:2005tq}. Pair production is also possible in the decay of nuclei. In this paper, we will derive an approximate result for $dN/dE_+$ of the produced positrons. The dynamics of nuclear fission are not well known \cite{bulgac}. We propose that the $e^+e^-$ pairs produced during fission can shed some light on the dynamics.

\begin{figure}

    \centering
    \includegraphics[width=0.5\textwidth]{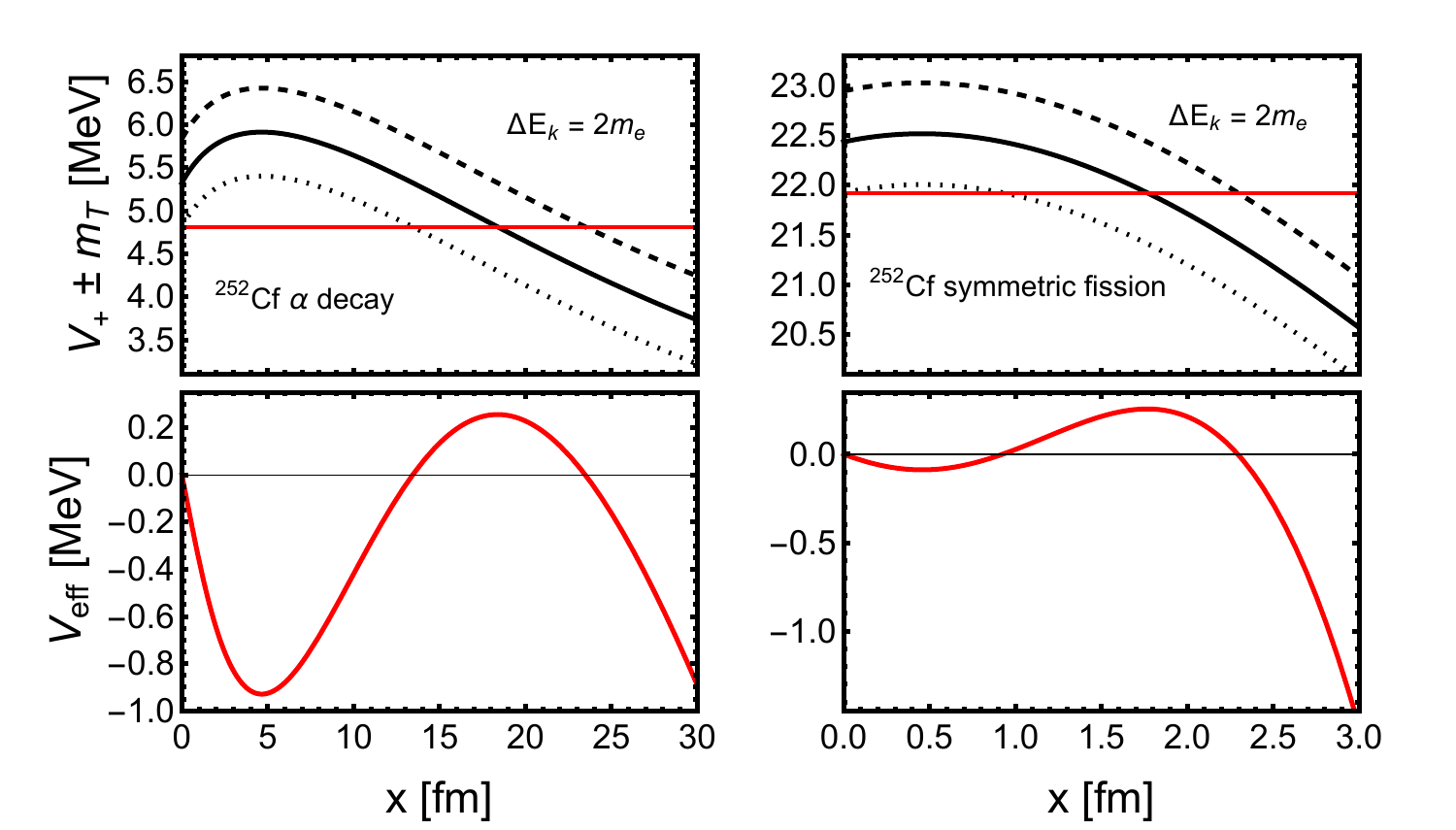}
    \caption{$V_+$ (top panels) and $V_{eff}$ (bottom panels) for $\alpha$ decay (left side) and symmetric fission (right side) of $^{252}$Cf. The relative distance between the ions is at the moment the $\alpha$ exits the Coulomb barrier in the case of $\alpha$ decay as dictated by the $Q$ value (6.2 MeV). For fission, the relative distance is when the two fission fragments are touching (12.0 fm).}
    \label{fig:veff}
\end{figure}

In our previous paper \cite{pairproduction}, we described a mechanism for pair production. We briefly recall the main points. In our geometry, two nuclei or fragments are a distance $R$ apart from each other. The electron is formed in the middle, and the positron tunnels on an $x$-axis perpendicular to the beam axis. The Coulomb potential seen by the positron is
\begin{equation}
    V_+(R, x) = \frac{Z_{tot} e^2}{\sqrt{ \left( \frac{R}{2} \right)^2+ x^2}} - S(x) \frac{e^2}{x}, \label{vc}
\end{equation}
where $x$ is the distance between the positron and electron, $Z_{tot} = Z_1 + Z_2$ is the total charge, $S(x) = 1 - \exp(-2m_T x / \hbar)$ is a screening factor, and $m_T = \sqrt{m^2+p_T^2}$ is the transverse mass of the positron. Ignoring small spin corrections, the positron satisfies the Klein-Gordon equation \cite{Schwinger:1951nm, wongbook} 
\begin{equation}
    \left[(E_+-V_+(R, x))^2-p_x^2-m_T^2\right]\psi = 0. \label{KG}
\end{equation}
This is formally equivalent to the Schr\"{o}dinger type equation
\begin{equation}
    \left[\frac{p_x^2}{2m_T} + \frac{m_T}{2} - \frac{(E_+ - V_+(R, x))^2}{2m_T}\right]\psi = 0,
\end{equation}
for a particle of mass $m_T$ in an effective potential
\begin{equation}
    V_{eff}(x) = \frac{m_T}{2} - \frac{(E_+ - V_+(R, x))^2}{2m_T} \label{veff}
\end{equation}
with effective energy $E_{eff}=0$. The effective potential has a hill that the positron can tunnel through (Fig. \ref{fig:veff}). The probability of tunneling is 
\begin{equation}
    \Pi_t = [1+\exp(2A/\hbar)]^{-1},
\end{equation}
where $A$ is the action integrated between the turning points.

In a small time $dt$, the probability of creating a pair is
\begin{equation}
    dN = \Pi_t \frac{dt}{\Delta \tau},
\end{equation}
where $\Delta \tau = \frac{\hbar}{2m_T}$ is the characteristic tunneling time from the Heisenberg principle \cite{pairproduction}. As $E_+$ approaches $m_T$, the barrier that the positron must tunnel through becomes infinitely long, so the tunneling probability goes to zero. From the conservation of energy, the energy associated with the positron is
\begin{equation}
    E_+ = \frac{2 Z_{tot} e^2}{R} - m_T - \Delta E_k, \label{E+}
\end{equation}
where $m_T$ is the transverse mass of the electron and $\Delta E_k$ is the change in kinetic energy of the ions at the moment of the pair's creation, which we showed must be $\geq 2m_T$ \cite{pairproduction}. A simple change of variables gives
\begin{equation}
    \frac{dN}{dE_+} = \frac{\Pi_t R^2}{2  \Delta \tau |\frac{dR}{dt}| Z_{tot} e^2}. \label{dnde}
\end{equation}
Thus the number of pairs produced at a certain energy depends on the distance between the ions and their relative velocity. Since the ions gain kinetic energy $\Delta E_k$ during the tunneling process, there is some ambiguity around which value of $dR/dt$ to use: the relative velocity before or after pair creation (or some average of the two). If we only consider the Coulomb repulsion between the ions, we can find $dR/dt$ from the Coulomb potential. Thus, for the Coulomb trajectory,
\begin{multline}
    \frac{dN}{dE_+} = \frac{\Pi_t \sqrt{2\mu} Z_{tot} e^2}{\Delta \tau (E_+ + m_T + \Delta E_k)^2} \\
    \times \left(E_{cm} + \beta \Delta E_k - \frac{Z_1 Z_2}{2Z_{tot}} (E_+ + m_T + \Delta E_k) \right)^{-1/2}, \label{onlycoulomb}
\end{multline}
where $\mu$ is the reduced mass of the ions, and $0 \leq \beta \leq 1$ is a parameter that determines the relative velocity of the ions based on the tunneling dynamics. In this way, pair production can give information about the tunneling process. 

\begin{figure}

    \centering
    \includegraphics{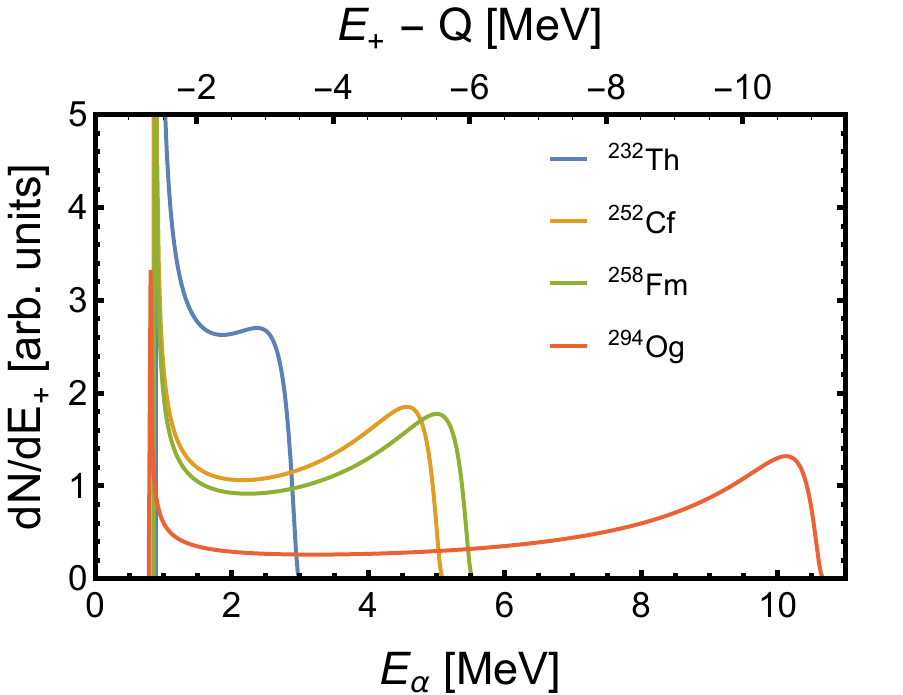}
    \caption{$dN/dE_+$ for $\alpha$ decay of a selection of nuclei. $\Delta E_k = 2m_e$ and $\beta = 0$ for this case.}
    \label{fig:alpha}
\end{figure}
We can apply this to the case of $\alpha$ decay. For this work, we assume that the $\alpha$ particle emerges from the Coulomb barrier at rest, and then it accelerates away from the daughter nucleus. Thus, Eq. \eqref{onlycoulomb} describes the distribution of positron energies for this case. As an approximation, we will say that the $\alpha$ particle takes away all the remaining kinetic energy so that
\begin{equation}
    E_\alpha = Q - E_+ - m_T.
\end{equation}
Figure \ref{fig:alpha} shows $dN/dE_+$ for the $\alpha$ decay of several heavy nuclei. The peaks on the left hand side of the graph correspond to positrons that are created when the $\alpha$ particle is at rest, since $|dR/dt|$ appears in the denominator of Eq. \eqref{dnde}. In reality, the $\alpha$ particle is not likely to be at rest when it emerges from the Coulomb barrier, so those peaks may not be reproduced in practice.


\begin{figure}
    \centering
    \includegraphics[width=0.5\textwidth]{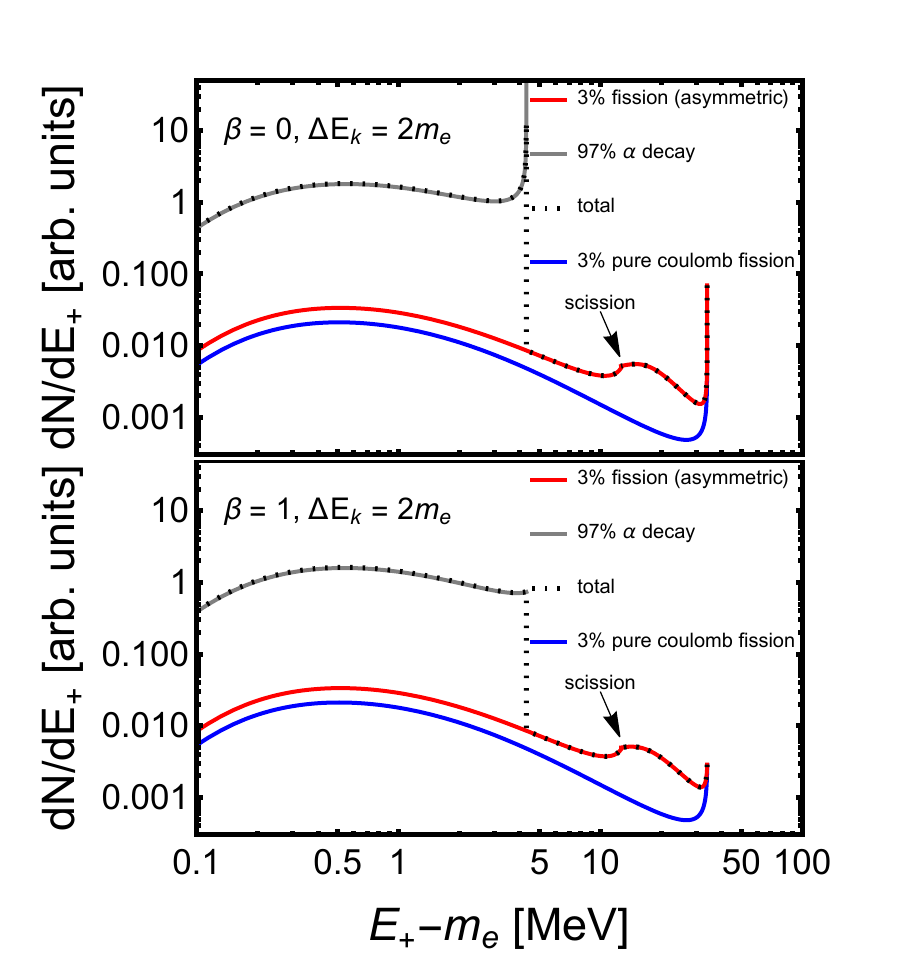}
    \caption{$dN/dE_+$ for fission and $\alpha$ decay of $^{252}$Cf, weighted by the corresponding branching ratios. Here, $m_T = m_e$ and $\Delta E_k = 2m_e$. In the top panel we assume that the extra acceleration ($\Delta E_k$) is given to the ions after the pair creation resulting in the divergence at time zero, $E_+\approx$ 30 MeV, $\beta=0$. The Coulomb divergence disappears if $\beta=1$, i.e. the ions are accelerated at the beginning of the pair production process (bottom panel). Similarly for $\alpha$ decay.}
    \label{fig:Cf}
\end{figure}

\begin{figure}
    \centering
    \includegraphics{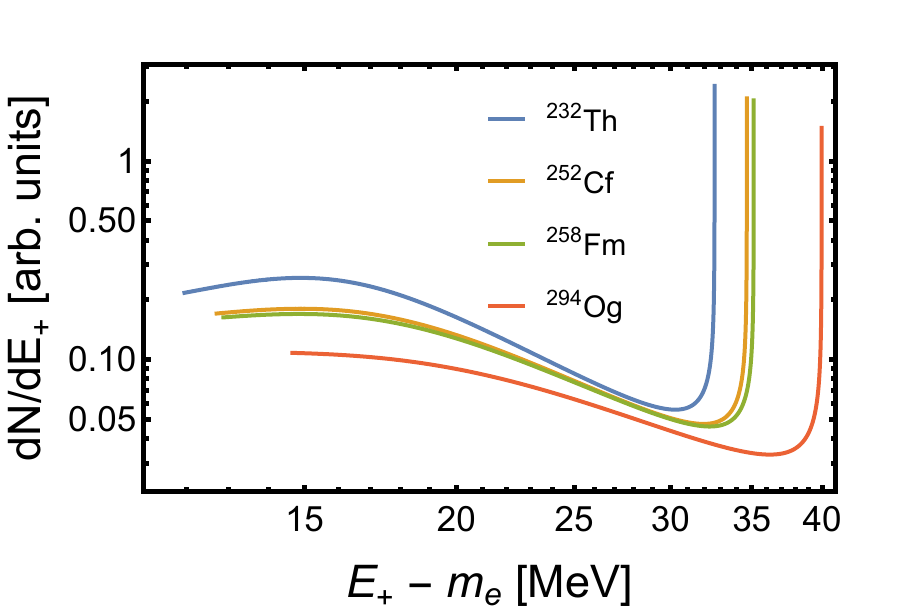}
    \caption{$dN/dE_+$ for fission of a selection of nuclei. The fission dynamics are based on the neck model. Here, $m_T = m_e$, $\beta=0$ and $\Delta E_k = 2m_e$.}
    \label{fig:fission}
\end{figure}

$dN/dE_+$ depends sensitively on the distance between the daughter nuclei at the moment the pair is created. Thus, information about the average dynamics of fission can be inferred from a careful measurement of the positron spectrum. A neck model \cite{neck} was used to simulate the dynamics for fission of several heavy nuclei. The neck model gives the distance between the fission fragments until the neck breaks and Coulomb repulsion takes over. 
During the fission process, the total nuclear volume of the fissioning fragments is assumed constant \cite{neckfiss}. The fragments separate because of the large Coulomb repulsion.  The separation is slowed down by the surface tension of the two fragments joined by a neck.  Furthermore, nucleon transfer through the neck results in the famous window dissipation formula \cite{randrup}.  These  ingredients make the dynamics up to the scission point quite different from the simple Coulomb repulsion of two point particles given by the full (blue) line in Fig. 3.  The pair production is strongly affected by the initial kinetic energy of the fission fragments. If the fragments start at rest, then a peak due to the Coulomb divergence, should be observed, Eq. \eqref{E+}. The peak disappears if in such a compact initial configuration some kinetic energy is initially given to the fragments.  This could be due to the action of the electron created between the two fragments as described above.
In Fig. \ref{fig:Cf}, $dN/dE_+$ for $\alpha$ decay and asymmetric fission of $^{252}$Cf \cite{neckfiss} is shown, multiplied by the corresponding branching ratios. The segment of the plot to the right of the point labelled ``scission" corresponds to pairs created during the fission process itself. As there are no positrons created at this high energy for $\alpha$ decay, we can see the effects of the fission dynamics. The blue curve shows the spectrum obtained if the fission fragments are allowed to accelerate away from each other without friction and just the Coulomb repulsion. The friction slows down the fission fragments, increasing $dN/dE_+$ for positrons of the appropriate energy. The positron spectra for symmetric fission of several heavy nuclei are shown in Fig. \ref{fig:fission}. Previous experimental investigations have looked for a coincidence between $e^+$ and $e^-$ \cite{tsunoda}. In our approach the coincidence is lost because the electron gets trapped by the nuclei. We suggest an experiment to look for $e^+$ in coincidence with a fission fragment and not with $e^-$.

In conclusion, our model predicts that electron-positron pairs can be created during $\alpha$ decay or fission. The energy of the created positron is related through the conservation of energy to the distance between the progeny nuclei. Careful observation of the energy spectrum of pairs produced during fission could reveal the dynamics of the fission process, properties of the vacuum polarization and the tunneling ``dynamics.''

\section*{Acknowledgements}
This research was funded in part by the United States Department of Energy under Grant \# DE-FG03-93ER40773 and the NNSA Grant No. DENA0003841 (CENTAUR) and by the National Natural Science Foundation of China (Grant Nos. 11905120 and 11947416).


\begin{thebibliography}{99}
\bibitem{Uehling:1935uj}
E.~A.~Uehling,
Polarization effects in the positron theory,
Phys. Rev. \textbf{48}, 55-63 (1935).

\bibitem{prestonbook}M. Preston, R. Bhaduri, Structure of the Nucleus, Addison-Wesley Publishing Company, Reading, MA, 1975.

\bibitem{Settlemyre:2021tbe}
T.~Settlemyre, H.~Zheng and A.~Bonasera,
Coulomb field correction due to virtual $e^+e^-$ production in heavy ion collisions,
Nucl. Phys. A \textbf{1015}, 122282 (2021)

\bibitem{assenbaum}
H.~Assenbaum, K.~Langanke and G.~Soff,
Vacuum polarization effects in subbarier nuclear fusion reactions,
Phys.~Lett.~B, \textbf{208} 346 (1988).

\bibitem{Schwinger:1951nm}
J.~S.~Schwinger,
On gauge invariance and vacuum polarization,
Phys. Rev. \textbf{82}, 664-679 (1951)

\bibitem{wongbook}C. Wong, Introduction to High-Energy Heavy-Ion Collisions, World Scientific Publishing, Singapore, 1994.

\bibitem{Blaschke:2019pnj}
D.~B.~Blaschke, L.~Juchnowski and A.~Otto,
Kinetic Approach to Pair Production in Strong Fields\textemdash{}Two Lessons for Applications to Heavy-Ion Collisions,
Particles \textbf{2}, no.2, 166-179 (2019)

\bibitem{carrollbook}B.~W.~Carroll and D.~A.~Ostlie, An Introduction to Modern Astrophysics, 2nd Edition, Addison-Wesley, San Francisco, 2007. 

\bibitem{Voskresensky:2021okp}
D.~N.~Voskresensky,
Electron-positron vacuum instability in strong electric fields. Relativistic semiclassical approach,
Universe \textbf{7}, no.4, 104 (2021)

\bibitem{popov}
R.~V.~Popov et al,
How to access QED at a supercritical Coulomb field.
Phys.~Rev.~D \textbf{102} 076005 (2020).

\bibitem{pairproduction} T.~Settlemyre, H.~Zheng and A.~Bonasera, Dynamical pair production at sub-barrer energies for light nuclei, Particles \textbf{5}, 580-588 (2022).

\bibitem{Kimura:2004dw}
S.~Kimura and A.~Bonasera,
Chaos Driven Fusion Enhancement Factor at Astrophysical Energies,
Phys. Rev. Lett. \textbf{93}, 262502 (2004)

\bibitem{Krauss:1987onj}
A.~Krauss, H.~W.~Becker, H.~P.~Trautvetter and C.~Rolfs,
Astrophysical S (E) factor of $^3$He($^3$He,2p)$^4$He at solar energies,
Nucl. Phys. A \textbf{467}, 273-290 (1987)

\bibitem{Engstler:1988tfw}
S.~Engstler, A.~Krauss, K.~Neldner, C.~Rolfs, U.~Schr\"oder and K.~Langanke,
Effects of electron screening on the $^3$He(d,p) $^4$He low-energy cross sections,
Phys. Lett. B \textbf{202}, 179-184 (1988)

\bibitem{Shoppa:1993zz}
T.~D.~Shoppa, S.~E.~Koonin, K.~Langanke and R.~Seki,
One- and two-electron atomic screening in fusion reactions,
Phys. Rev. C \textbf{48}, 837-840 (1993)

\bibitem{Musumarra:2001xd}
A.~Musumarra, R.~G.~Pizzone, S.~Blagus, M.~Bogovac, P.~Figuera, M.~Lattuada, M.~Milin, \DH{}.~Miljanic, M.~G.~Pellegriti and D.~Rendic, \textit{et al.}
Improved information on the $^2$H($^6$Li,$\alpha$)$^4$He reaction extracted via the Trojan Horse Method,
Phys. Rev. C \textbf{64}, 068801 (2001)

\bibitem{Kimura:2005tq}
S.~Kimura, A.~Bonasera and S.~Cavallaro,
Influence of chaos on the fusion enhancement by electron screening,
Nucl. Phys. A \textbf{759}, 229-244 (2005)

\bibitem{bulgac}  Bulgac A, Jin S and Stetcu I Nuclear Fission Dynamics: Past, Present, Needs, and Future. Front. Phys. \textbf{8} 63 (2020). 

\bibitem{neck} A.~ Bonasera, G.~Bertsch and E.~El Sayed, Newtonian dynamics of time-dependent mean field theory, Phys. Lett. B \textbf{141} 9-13 (1984).

\bibitem{neckfiss} A.~Bonasera, Dynamical model of nuclear fission with shell effects, Phys. Rev. C \textbf{34} 740-742 (1986).

\bibitem{randrup} J.~Randrup and W.~J.~Swiatecki. Dissipative resistance against changes in the mass asymmetry degree of freedom in nuclear dynamics: The completed wall-and-window formula, Nucl. Phys. A \textbf{429} 105-115 (1984).

\bibitem{tsunoda}
T.~Tsunoda et al,
A search for correlated $e^+e^-$ pairs in the fission process,
Workshop on dynamical symmetry breaking 175-180 (1989).

\end{thebibliography}
\end{document}